\documentstyle[12pt,a4,epsfig]{article}
\def\g5{\gamma_5}

\begin{document}
\sloppy
\thispagestyle{empty}

\mbox{}
%%%%%%%\vspace*{\fill}

\rightline{MSU-51203}

\begin{center}
{\LARGE\bf Single Spin Asymmetries at RHIC\footnote{Talk presented
during the meeting on {\it 'The Measurement of Singly Polarized
Collisions at HERA'} held at DESY-Zeuthen, Germany, 31 August-1
September 1995.}} \\

\vspace{2em}
\large
Glenn A. Ladinsky
\\
\vspace{2em}
{\it  Michigan State University}
 \\
{\it Department of Physics \& Astronomy, East Lansing, MI  48824-1116
 U.S.A.}\\
%\today
\end{center}
%%%%%%%\vspace*{\fill}
%
\begin{abstract}
\noindent
One purpose of this meeting is to assess the physics potential
of HERA-N Stage I.  To develop a reasonable perspective,
it is useful to look at other spin experiments. For completeness, 
this report discusses some of the single spin physics that has appeared
in studies relating to the Relativistic Heavy Ion Collider (RHIC).
\end{abstract}

%%%I don't see a need to start a new page.

%\vspace*{\fill}
%\newpage

\vspace{1.0cm}

\def\greatvoid#1{ }
\def\journal#1&#2(#3)#4{{\unskip,~\sl #1\unskip~\bf\ignorespaces #2\unskip~\rm (19#3) #4}}
\def\jour#1&#2(#3)#4{{\unskip ~\sl #1\unskip~\bf\ignorespaces #2\unskip~\rm (19#3) #4}}\def\p{{\bf p}}

\section{Introduction}
\label{sect1}

One purpose of this meeting is to assess the physics potential
of HERA-N Stage-I (HNSI).  To develop a reasonable perspective,
it is useful to look at other spin experiments. For completeness, 
this report discusses some of the single spin physics that has appeared
in studies relating to the Relativistic Heavy Ion Collider (RHIC).

Recall\cite{nowak}, in HNSI an unpolarized proton beam of
$820\,$GeV would collide with a polarized nuclear target enabling us
to study proton-proton and proton-neutron asymmetries at a center of
mass energy near $\sqrt{S}\approx 40\,$GeV.  The luminosities
anticipated range is from $8-240\,\hbox{pb}^{-1}$.  For understanding
what physics can be expected from this single spin experiment, we can
ask about the relevance past studies performed for RHIC 
have with regard to HNSI.

It is important to note that there are significant
differences between HNSI and the early runs at RHIC.
The spin program at RHIC\cite{prop,robin} involves collisions between
two polarized proton beams.  The energy range for the early runs 
are expected to be from $200-500\,$GeV, though in the end the range
may run from $50-500\,$GeV.  The integrated luminosity expected is about
$800\hbox{pb}^{-1}$ ($320\hbox{pb}^{-1}$) at $\sqrt{S}=500\,$GeV ($50\,$GeV)
with beam polarizations roughly around 70\%.

This disparity between the two machines leads to a tremendous difference 
in the focus between the RHIC project and HSNI.
In the first place, both beam spins will be under control at RHIC, 
so their focus is naturally on double spin asymmetries.  
Single spin asymmetries from RHIC can be obtained
for free, however, simply by combining cross sections with complementary
spin combinations.  In the second place, the RHIC energies are
high enough to study $W^\pm$ and $Z^0$ boson physics.  Parity violation
is manifest in the weak interactions and may even be used as a tool.

In the original RHIC spin 
proposal little is said about single spin asymmetries\cite{prop}.
One item refers to the potential of measuring the longitudinal
single spin asymmetry $A_L$ and the transverse spin asymmetry
$A_N$ in elastic scattering at small $|t|$.
What is also needed is a study of the energy dependence
of the forward hadron reactions in the total cross section,
\begin{eqnarray}
\sigma^{tot}&=&{2\pi\over k} \hbox{Im}[\phi_1(0)+\phi_3(0)],\\
\Delta\sigma_{L}&=&{4\pi\over k} \hbox{Im}[\phi_1(0)-\phi_3(0)],\\
\Delta\sigma_{T}&=&-{4\pi\over k} \hbox{Im}[\phi_2(0)],
\end{eqnarray}
where $\phi_{1,2,3}$ are the $s$-channel helicity amplitudes.

Also discussed is how measuring $A_N$ (the single spin transverse
asymmetry) in 
$$
p^\uparrow p\rightarrow \gamma+X
\qquad \hbox{and} \qquad
p^\uparrow p\rightarrow \pi^0+X
$$
will probe the twist-3 structure of the hadron.
Since higher twists should not contribute at high
energies, one could also test the prediction of perturbative QCD
for a vanishing
one-spin transverse asymmetry ($A_N=0$) in high-$P_{T}$ inclusive
production.  The study of the $P_{T}$ dependence of one-spin
asymmetries might reveal the transition from the nonperturbative phase
of QCD ($A_N\ne 0$) to the perturbative part ($A_N=0$).
Moreover, this experiment is clean compared to longitudinal spin
and double spin experiments since it doesn't have leading twist variations
tangled together with the effects due to nonleading twist.
Such data also will provide information on the correlations between 
the quarks and gluons
inside the proton, which can be used to constrain the various nucleon models.

In QCD, where parity is conserved, the single spin longitudinal asymmetry
($A_L$) is zero at the tree level.  In contradistinction to this,
the weak interactions typically yield $A_L\ne 0$.  Looking at the
factorization of the cross section for $A_L$,
\begin{equation}
A_L \propto \sum_{\hbox{partons}\, a,b} 
\int dx_1\, dx_2\, \Delta f_{a/A}(x_1,Q)
f_{b/B}(x_2,Q)\, \hat{a} \, d\sigma,
\end{equation}
it is apparent that the nonzero asymmetry of the hard interaction,
$\hat{a}$, permits a probe of the helicity distribution
$\Delta f(x,Q)$.  
In Fig.~1, it is depicted how the 
production of gauge bosons will probe the helicity distributions for sea
quarks through $A_L$.  

\vbox{
\epsfysize=3.0in
\leftline{\epsfbox{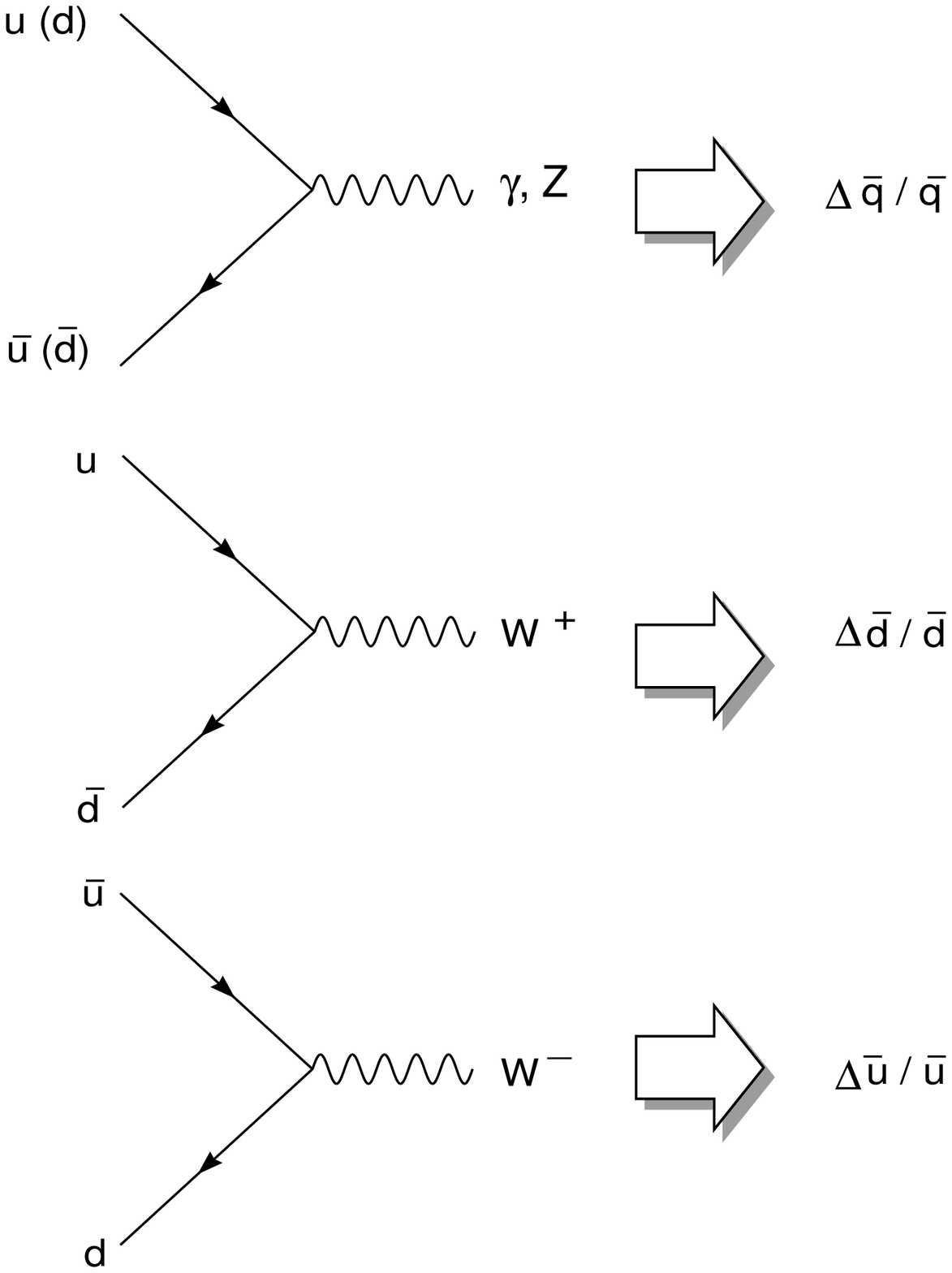}}

\vspace{-1.75in}
\rightline{
\parbox{3.5in}{Fig.~1:  Different flavors for the helicity densities
of the sea quarks are probed depending upon which gauge boson is
produced.}
}
\vspace{+1.25in}
}

The single spin asymmetries have been computed for the Drell-Yan process
and for gauge boson production.  In particular, Leader and Sridhar have
computed $A_L$ for the Drell-Yan process including the interference
effects between the photon and $Z$ boson at $\sqrt{S}=500\,$GeV.  
As would be expected, the parity violating
asymmetry is most sensitive to variations in the helicity
densities as the energy for the subprocess approaches the $Z$ boson
mass.  Variations on the order of 20\% may appear.

Bourrely and Soffer have computed the asymmetries for electroweak
boson production at $\sqrt{S}=500\,$GeV\cite{sofw}.  They find 
the single spin asymmetry spanning ranges as large as 60\%
across the boson's rapidity.  The largest variation with a change
in polarized parton distributions appears for the $W^+$ boson.
This should provide a decent probe of the $\Delta\bar{d}$
distribution function.  In fact, studies by the STAR and
PHENIX Collaborations indicate that with the tens of thousands
of $W$ boson events they can collect, the statistical bound on
how well we can determine 
$\Delta \bar{u}/\bar{u}$ and $\Delta \bar{d}/\bar{d}$ is 
to within 3\%\cite{review}.

Jet production, even though dominated by QCD processes, may carry a parity
violating behavior due to the electroweak production mechanisms and
their interference with QCD processes\cite{ranft}.
This parity violation can appear in $A_L$.
Without cuts, Ref.~\cite{ranft} demonstrates
asymmetries around the percent level for $pp$ and $p\bar{p}$
collisions ranging from $\sqrt{s}=250-850\,$GeV.
With high luminosities, the RHIC
should be able to observe some variation.

Though parity conservation yields $A_L=0$ for single particle
inclusive processes at the tree level, 
it is possible to get nonzero asymmetries by observing more than
one particle in the final state.  In particular, Pire and Ralston\cite{pire}
and Carlitz and Willey\cite{willey} demonstrate nonzero asymmetries that
appear at the one loop level in the production of dilepton pairs.
These dileptons may be produced by a Drell-Yan mechanism or through
decays {\it e.g.}, of $J/\psi$.

The longitudinal asymmetry arises from the vector product 
${\bf s\cdot (q^+\times q^-)}$, which is odd under time reversal.
(${\bf s}$ is the spin vector, ${\bf q^+}$ is the $\mu^+$ momentum and 
${\bf q^-}$ is the $\mu^-$ momentum.)
Consequently, imaginary parts of the one loop amplitude in the lepton
pair production provide for the nonzero asymmetry.  The parton level
asymmetry corresponding to $A_L$ is computed by Carlitz and Willey and
found to $10-30\%$ for $\tau=-t/({\bf q^+}+{\bf q^-})^2<10$.
Ideally, by using different kinematics, it should be possible to focus on
subprocesses with either the quarks or gluons in the initial state.
Through this asymmetry, it becomes possible to collect information
on the helicity distributions.
These ideas can be applied to other processes also, like open heavy quark
production, dijet production and $\gamma+$jet production.
At RHIC, however, such data cannot compete with the probe into
the parton distribution functions that the double spin asymmetries
can provide.

Other opportunities exist.  For example, in the forward scattering region
the outgoing quark is expected to maintain the helicity of its original
state.  Analyzing the polarized fragmentation of the resulting particle
jet can provide information on the the chiral structure of QCD
and also yield a useful tool for further polarization 
experiments\cite{jccfrg}.

It was not the intent of this talk to enter into a description
of the theoretical details regarding single spin asymmetries.
For further discussion on theoretical motivations for
studying single spin physics at hadron-hadron colliders, 
I direct the reader to the 
{\it Review of the RHIC Spin Physics Program}\cite{review}
and to an excellent
review written by S.M.~Troshin and A.~Krisch
in the {\it Acceleration of Polarized Protons to 120 GeV and 1 TeV at Fermilab}
for the SPIN Collaboration.  Though the discussion
in the latter document is directed toward the Fermilab facility, 
many of the ideas
apply equally well to the high energy collisions at RHIC.

\section*{Acknowledgments}
\label{sect6}

My gratitude goes to Wolf-Dieter Nowak and Johannes Blumlein
for inviting me to DESY-Zeuthen and for their hospitality 
during my stay.  For their useful comments and discussion I also
thank G.~Bunce, J.C.~Collins, E.~Berger, G.~Sterman
and M.~Tannenbaum.
This work is funded in part by DESY-Zeuthen,
Michigan State University and NSF grant PHY-9309902.

%%%%%%%%%%%%%%%%%%%%%%%%%%%%%%%%%%%%%%%%%%%%%%%%%%%%%%%%%%%%%%%%%%%%%%%%%%
%%References start:
%%%%%%%%%%%%%%%%%%%%%%%%%%%%%%%%%%%%%%%%%%%%%%%%%%%%%%%%%%%%%%%%%%%%%%%%%%

\newpage
%%%%%%%%%%%%%%%%%%%%%%%%%%%%%%%%%%%% figures %%%%%%%%%%%%%%%%%%%%%%%%%%%
%%\newpage

%\begin{center}
%\mbox{\epsfig{file=fig1.eps,height=18cm,width=18cm}}
%\vspace{2mm}
%\noindent
%\small
%\end{center}
%{\sf Figure~1:}~My figure 1.
%\normalsize

\end{document}